\documentclass[aip,preprint]{revtex4}

\usepackage{color}
\usepackage{amsmath}
\usepackage{amsfonts}
\usepackage{graphicx}

\begin{document}

\title{On optical spectroscopy of molecular junctions}

\author{Michael Galperin}
\affiliation{Department of Chemistry \& Biochemistry, University of California San Diego, La Jolla, CA 92093, USA}
\author{Mark A. Ratner}
\affiliation{Department of Chemistry, Northwestern University, Evanston, IL 60208, USA}
\author{Abraham Nitzan}
\affiliation{School of Chemistry, Tel Aviv University, Tel Aviv, 69978, Israel}

\date{\today}

\begin{abstract}
We compare theoretical techniques utilized for description of 
optical response in molecular junctions, and their application 
to simulate Raman spectroscopy in such systems.
Strong and weak sides of the Hilbert vs. Liouville space, 
as well as quasiparticles vs. many-body states, formulations are discussed.
Common origins of the methodologies and different approximations
utilized in different formulations are identified.
\end{abstract}

\maketitle

\section{Introduction}\label{intro}
The interaction of light with molecules is an important field of research due to
its ability to provide information on molecular structure and dynamics,
and to serve as a control tool for intra-molecular processes. 
Theory of molecular optical spectroscopy has been developed and widely utilized 
in studies of optical response of molecules either in the gas phase or 
chemisorbed on surfaces~\cite{Mukamel_1995}. 

Recent progress in nano fabrication made it possible to perform optical experiments on molecular conduction junctions. 
In particular, current induced fluorescence\cite{HoPRB08} and Raman 
measurements\cite{CheshnovskySelzerNatNano08,NatelsonNL08,NatelsonNatNano11} were reported in the literature. Theoretical description of optical response of such
nonequilibrium open molecular systems is challenging
due to necessity to account for optical excitations in the system in the presence of 
electron flux through the junction. For example, correlation between
Stokes signal and conductance were measured\cite{NatelsonNL08} 
and an attempt of theoretical explanation op the effect was proposed\cite{ParkGalperinEPL11,ParkGalperinPRB11}.
With scattering theory being inapplicable to
the many-body electronic problem of quantum transport in junctions, 
the very definition of optical scattering processes in such systems is a challenge.
Also, as discussed below, formal application of the molecular spectroscopy theory to 
junctions may be problematic.

Here we compare theoretical approaches utilized in the literature
for description of optical spectroscopy of current carrying molecular junction. 
After introducing model of the junction
subjected to external radiation field in Section~\ref{model},
we consider theoretical foundations of
molecular spectroscopy in Section~\ref{phflux}. We then discuss
Hilbert vs. Liouville space formulations in 
Section~\ref{hilbert_liouv} and
quasiparticles vs. many-body states versions of the two formulations
in Section~\ref{qp_mb}. In Section~\ref{raman} we specifically focus
on different contributions to the total optical signal, and
identify those relevant for the Raman scattering.
Our conclusions are presented in Section~\ref{conclude}.


\section{Model}\label{model}
We consider a junction formed by a molecule $M$ coupled to two contacts $L$ and $R$
(each at its own thermal equilibrium) subjected to an external radiation field.
Hamiltonian of the model is
\begin{align}
\label{H}
\hat H =& \hat H_0 + \hat V
\\
\label{H0}
\hat H_0 =& \hat H_S + \sum_{B=L,R}\bigg(\hat H_B+\hat V_{SB}\bigg) + \hat H_{rad}
\\
\label{V}
\hat V =&  \sum_{M,\alpha}\left(U_{M,\alpha}\hat O_M^\dagger\hat a_\alpha
 + H.c.  \right)  
\end{align}
where $\hat H_0$ describes molecular junction and radiation field, and
$\hat V$ introduces coupling between them. 
Here $\hat H_S$, $\hat H_B$, and $\hat H_{rad}$ are Hamiltonians
of the molecule (system), contacts (baths), and radiation field, respectively. 
$\hat H_S$ describes electronic and vibrational structure of the molecule
as well as arbitrary intra-molecular interactions.
The contacts and field are assumed to be reservoirs of free electrons and photons, respectively
\begin{align}
\hat H_B =& \sum_{k\in B} \varepsilon_k\hat c_k^\dagger\hat c_k
\\
\hat H_{rad} =& \sum_\alpha \nu_\alpha \hat a_\alpha^\dagger\hat a_\alpha
\end{align}
$\hat V_{SB}$ couples molecule with contacts and usually is assumed to be 
quadratic in the quasiparticle representation
\begin{equation}
\label{VSB}
 \hat V_{SB} = \sum_{m\in S,k\in B} \bigg( V_{km}\hat c_k^\dagger\hat d_m
 + V_{mk} \hat d_m^\dagger \hat c_k\bigg) 
\end{equation}
In Eqs.~(\ref{H})-(\ref{VSB}) $\hat d_m^\dagger$ ($\hat d_m$) and $\hat c_k^\dagger$ ($\hat c_k$) are the creation (annihilation) operators of electron in level $m$ of the molecule
and state $k$ of the contact, respectively. $\hat a_\alpha^\dagger$ ($\hat a_\alpha$)
creates (destroys) photon in mode $\alpha$ of the radiation field, 
$\hat O_M^\dagger$ ($\hat O_M$) is operator creating (destroying) molecular optical
excitation $M$, 
and $U_{M,\alpha}$ is coupling between the excitation $M$ 
and mode $\alpha$ of the radiation field. Here and below $\hbar=e=1$.

\begin{figure}[t]
\centering\includegraphics[width=0.8\linewidth]{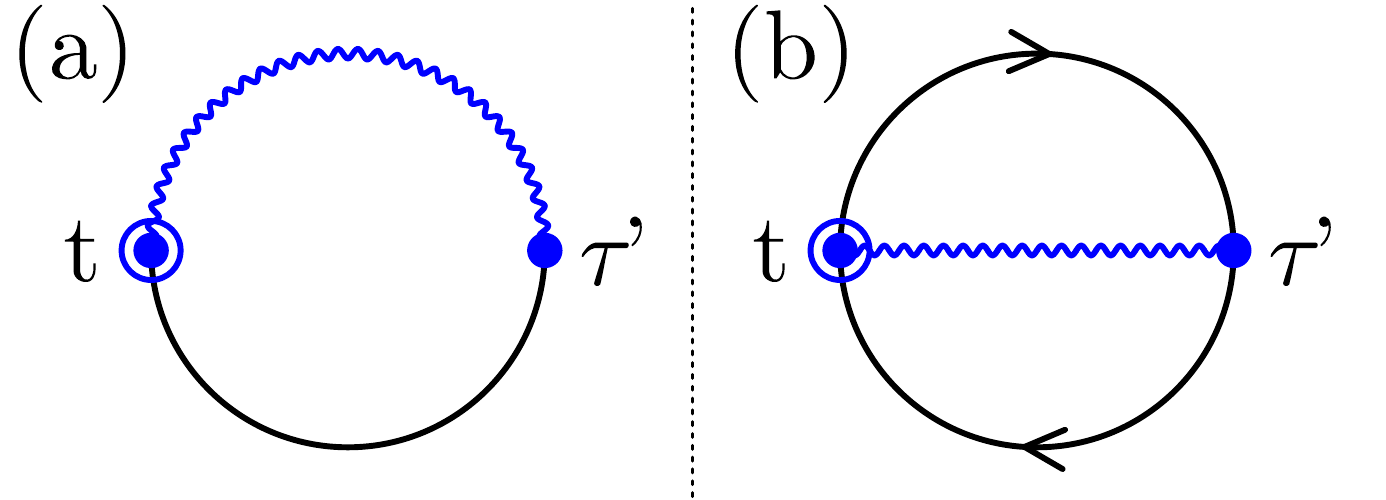}
\caption{\label{fig1}
Photon flux diagrams in
(a) quasiparticle and (b) many-body states representations.
Solid and wavy lines represent electron and photon propagators, respectively.
Non-directed lines indicate both possible directions. $t$ is the time of the flux,
$\tau'$ is the Keldysh contour integration variable. 
Summation over all indices and integration over contour variables is assumed for every connection in the diagrams (except the fixed time of the flux $t$). 
}
\end{figure}
\section{Optical signal as photon flux}\label{phflux}
Similar to considerations of electron transport in junctions, 
theories of optical spectroscopy are mostly focused on simulating fluxes.
More advanced treatments involve also higher cumulants of
the counting statistics~\cite{EspositoRMP09}. 
Diagrams for photon (boson) fluxes are presented in Fig.~\ref{fig1}.
As usual flux is defined as rate of population change in the continuum of modes $\{\alpha\}$ of the radiation field, 
and corresponding expression for the photon flux from the system 
into modes of the field at time $t$
is~\cite{GalperinNitzanRatner_heat_PRB07,GalperinRatnerNitzanJCP09}
\begin{align}
\label{flux}
 J(t) \equiv& \sum_\alpha\frac{d}{dt}\langle\hat a_\alpha^\dagger(t)\hat a_\alpha(t)\rangle
\\
 =& 2\mbox{Re}\int_{-\infty}^t dt'\,
 \mbox{Tr}\bigg[\mathcal{G}^{>}(t,t')\,\Pi^{<}(t',t)\,
 -\mathcal{G}^{<}(t,t')\,\Pi^{>}(t',t) \bigg]
\nonumber
\end{align}
Here $\mbox{Tr}[\ldots]$ is trace over molecular optical excitations $M$
(depending on the formulation these are either of quasiparticle type 
or represent transitions between many-body states of the molecule), 
$\Pi^{<(>)}$ and $\mathcal{G}^{<(>)}$ are lesser (greater) 
projections of the self-energy due to coupling to the radiation field
and correlation (Green) function of molecular optical excitations,
respectively~\cite{HaugJauho_2008}
\begin{align}
 \label{defPI}
 \Pi_{M_1,M_2}(\tau',\tau) \equiv& 
 \sum_{\alpha} U_{M_1,\alpha}\, F_\alpha(\tau',\tau)\, U_{\alpha,M_2}
 \\
 \label{defG}
 \mathcal{G}_{M_1,M_2}(\tau,\tau') \equiv&
 -i\langle T_c\, \hat O_{M_1}(\tau)\, \hat O^\dagger_{M_2}(\tau')\rangle
\end{align}
where $\tau$ and $\tau'$ are the Keldysh 
contour variables corresponding to real times $t$ and $t'$, 
$T_c$ is the contour ordering operator, and
\begin{equation}
 F_\alpha(\tau',\tau) \equiv 
 -i\langle T_c\, \hat a_\alpha(\tau')\, \hat a^\dagger_\alpha(\tau)\rangle
\end{equation}
is the Green function of free photon evolution. 
Its lesser and greater projections are
\begin{align}
& F^{<}_\alpha(t',t) = -i N_\alpha e^{-i\nu_\alpha(t'-t)}
 \\
 & F^{>}_\alpha(t',t) = -i [1+N_\alpha] e^{-i\nu_\alpha(t'-t)}
\end{align}
where $N_\alpha$ is average population of the mode $\alpha$.

We note that expression~(\ref{flux}) is exact. It consists of two
contributions: in-scattering (photon absorbed by the system;
first term in the second row) and out-scattering (photon emitted by the
system; second term in the second row). The former characterizes e.g.
absorption spectrum, while the latter can give information on 
fluorescence. Standard treatment proceeds by evaluating the 
molecular correlation functions $\mathcal{G}^{<(>)}$ in the presence of
intra-molecular interactions, coupling to environment
(e.g. molecular coupling to contacts in junctions),
and radiation field. Depending on the formulation some or all
of these processes are taken into account in the molecular Green function
through corresponding self-energies $\Sigma$. 
The latter usually can be evaluated
only approximately. Corresponding expressions should be derived from the
Luttinger-Ward functional~\cite{LuttingerWardPR60}, 
so that the resulting approximation fulfils conservation 
laws~\cite{KadanoffBaym_1962,vanLeeuwenPRB09}. 
The molecular Green function can be evaluated utilizing the Dyson equation
\begin{equation}
\label{Dyson}
 \mathcal{G}(\tau,\tau') = \mathcal{G}_0(\tau,\tau')
 + \int_c d\tau_1\int_c d\tau_2\, \mathcal{G}_0(\tau,\tau_1)\,
 \Sigma(\tau_1,\tau_2)\, \mathcal{G}(\tau_2,\tau')
\end{equation}
where $\mathcal{G}_0$ is the Green function in the absence of interactions.
In most cases Eq.~(\ref{Dyson}) has to be solved self-consistently due to dependence 
of the self-energy $\Sigma$ on the molecular Green function.
Substituting the converged result into (\ref{flux}) yields information
on the incoming, outgoing, or total optical flux. 

In practice treatment of molecular correlation function is often done
in a simplified manner directly employing perturbation theory
(at least in coupling to the radiation field).
While this way conserving character of the approximation
cannot be guaranteed (e.g. lowest order in electron-photon
interaction - the Born approximation - is known 
to be non-conserving
and in some cases may be not fully adequate~\cite{ParkGalperin_FCS_PRB11}),
the simplification avoids necessity of self-consistent 
treatment, and allows direct classification of multi-photon processes
in the system. For example, expanding the Green function 
$\mathcal{G}$, Eq.~(\ref{defG}), in pertubation series up to the second order in 
interaction with the radiation field $\hat V$, Eq.~(\ref{V}), yields
\begin{equation}
 \label{PT}
 \mathcal{G}(\tau,\tau') = \mathcal{G}_0(\tau,\tau')
 -\int_c d\tau_1\int_c d\tau_2\, \Pi(\tau_1,\tau_2)\,
 \langle T_c\, \hat O(\tau)\,\hat O^\dagger(\tau_1)\,
 \hat O(\tau_2)\, \hat O^\dagger(\tau') \rangle_0 
\end{equation}
where subscript $0$ indicates that corresponding averages 
are evaluated in the absence of the field (intra-molecular interactions and coupling to
contacts are still present).
Substituting first term in the right into (\ref{flux})
yields lowest (second) order contributions into absorption (in-scattering)
and emission (out-scattering) spectrum. 
Substitution of the second term in the right provides information 
on all fourth order optical processes in the system.

Below we focus on the out-scattering photon flux, second term in
the bottom row of Eq.~(\ref{flux}), and discuss fourth order optical
processes, second term in the right of Eq.~(\ref{PT}), 
in the emission spectrum within the Hilbert and Lioville space formulations.
Following the tradition we pick one of the modes 
(or set of modes at particular frequency) in the sum (\ref{defPI}),
substitute it in place of the self-energy in second term in the bottom row
of Eq.~(\ref{flux}), and assume the mode (which we call final, $f$)
to be empty. Similarly, in Eq.~(\ref{PT}) we pick one of the modes
(which we call initial, $i$) from the sum and use it in place of the
self-energy in the expression. This leads to
\begin{align}
 \label{Jif}
 J_{i\to f}(t) =& 2\,\mbox{Re} \int_{-\infty}^t dt' \int_c d\tau_1\int_c d\tau_2
 \sum_{M,M',M_1,M_2} U_{M',f}U_{f,M}U_{M_1,i}U_{i,M_2} 
 F_f^{>}(t'-t)F_i(\tau_1,\tau_2)
 \nonumber \\ &
 \langle T_c\, \hat O_{M'}^\dagger(t')\, \hat O_M(t)\,
 \hat O_{M_1}^\dagger(\tau_1)\, \hat O_{M_2}(\tau_2) \rangle
\end{align}
We note in passing that separation of the self-energies $\Pi$
into modes in principle can be avoided 
(see e.g. Refs.~\onlinecite{WhiteFainbergGalperinJPCL12,BaratzWhiteMGBaerJPCL14}).


\section{Hilbert vs. Liouville space formulation}\label{hilbert_liouv}
Difference between Hilbert vs. Liouville space formulations stems from
the way projections of contour variables are performed. 
The Hilbert space formulation utilizes the Keldysh contour, and thus transition
from contour variables to real times (i.e. ordering positions of the times
on the contour) requires to work with several types of 
projected correlation functions
(depending on positions of the times on the contour).
The Liouville space formulation works with the real time axis, ordering
the times relative to each other on the axis, and thus necessitates
introduction of Liouville superoperators, whose role
- distinguish between action on different branches of the contour -  
is exactly the same as that of different correlation functions projections
in the Hilbert space formulation. Naturally, both formulations yield the same result~\cite{SchwingerJMP61,KeldyshSovPhysJETP65,CraigJMP68,SchonPRB94,Harbola2006c,HarbolaMukamelPhysRep08}.

\begin{figure}[t]
\centering\includegraphics[width=0.8\linewidth]{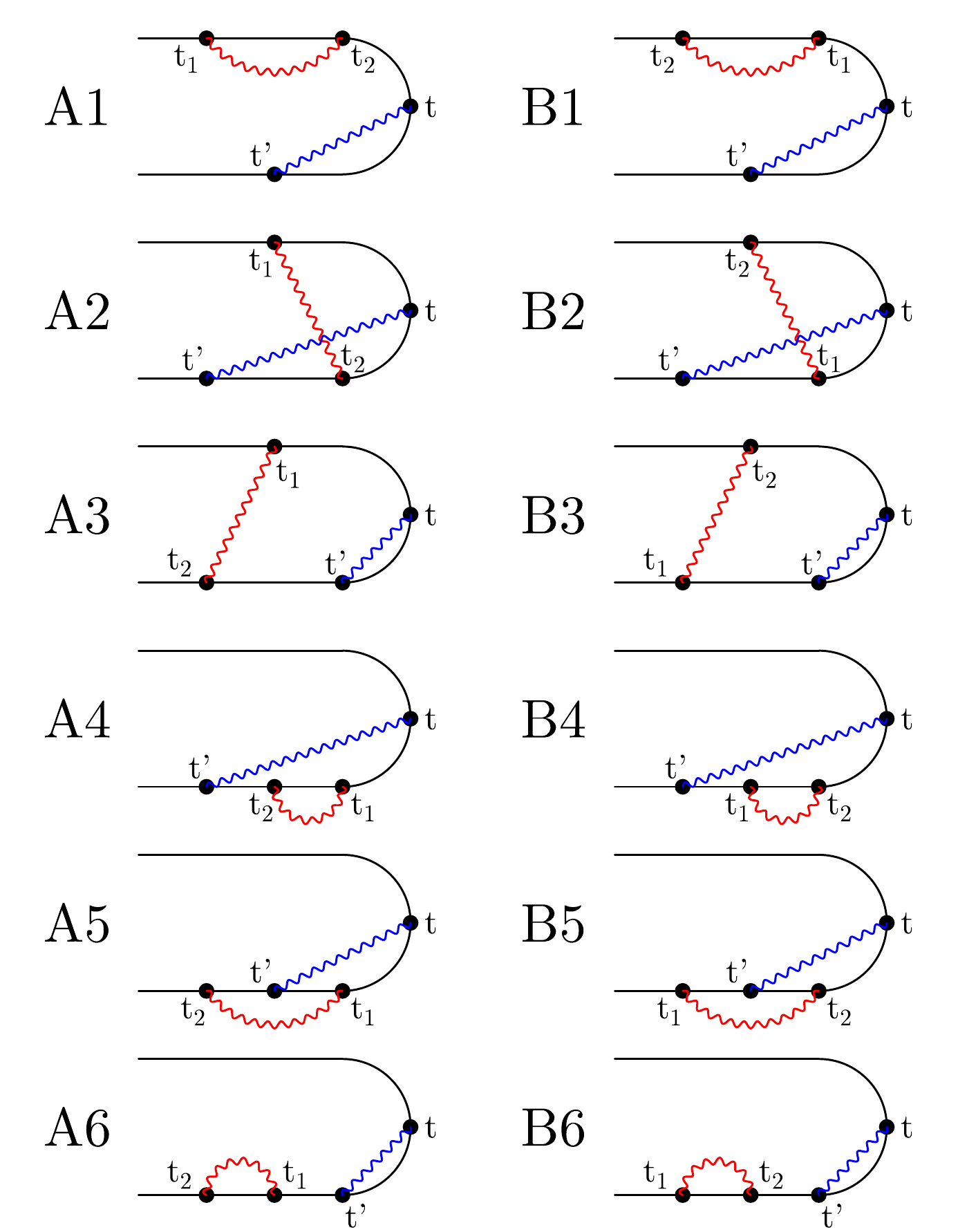}
\caption{\label{fig2}
Contour projections for the fourth order interaction with the modes 
$i$ (red, times $t_1$ and $t_2$) and $f$ (blue, times $t$ and $t'$) 
of the optical field. Time increases from left to right in these diagrams.
}
\end{figure}

Utilizing the Hilbert space procedure in Eq.~(\ref{Jif})
is equivalent to consideration of all possible placements of
times $t_1$ and $t_2$ (real times corresponding to contour
variables $\tau_1$ and $\tau_2$) between times $t$ and $t'$ on the Keldysh contour.
Note that in Eq.~(\ref{Jif}) the latter are placed 
in such a way that $t'$ follows $t$ on the contour
(this results form the greater character of the photon 
Green function $F_f^{>}(t',t)$). These placements lead to
$12$ possible diagrams (i.e. time orderings on the contour)
shown in Fig.~\ref{fig2}. One can also get $12$ additional diagrams,
which are just complex conjugate versions of those presented in Fig.~\ref{fig2}
(they come from the $\mbox{Re}$ in Eq.~(\ref{Jif})).
Causal character of interactions implies that time $t$ (the time of the signal)
is the latest time, usually placed at the point where the two 
branches of the contour meet each other. $t'$ then belongs to the 
anti-ordering branch, and $t_1$ and $t_2$ may populate any of the three
regions (preceding $t$, between $t$ and $t'$ and after $t'$ on the contour).
Also times $t$ and $t'$ as well $t_1$ and $t_2$ are connected by 
lines representing free photon propagation described by Green functions
$F$. Character of the Green functions (greater for the pair $t$ and $t'$,
$F_f^{>}(t',t)$, and either greater or lesser for the pair $t_1$ and $t_2$,
$F_i^{>(<)}(t_1,t_2)$) depends on ordering of the times on the contour
and represents photon emitted (for the greater projection, out-scattering)
or absorbed (for the lesser projection, in-scattering) by the system.

\begin{figure}[t]
\centering\includegraphics[width=\linewidth]{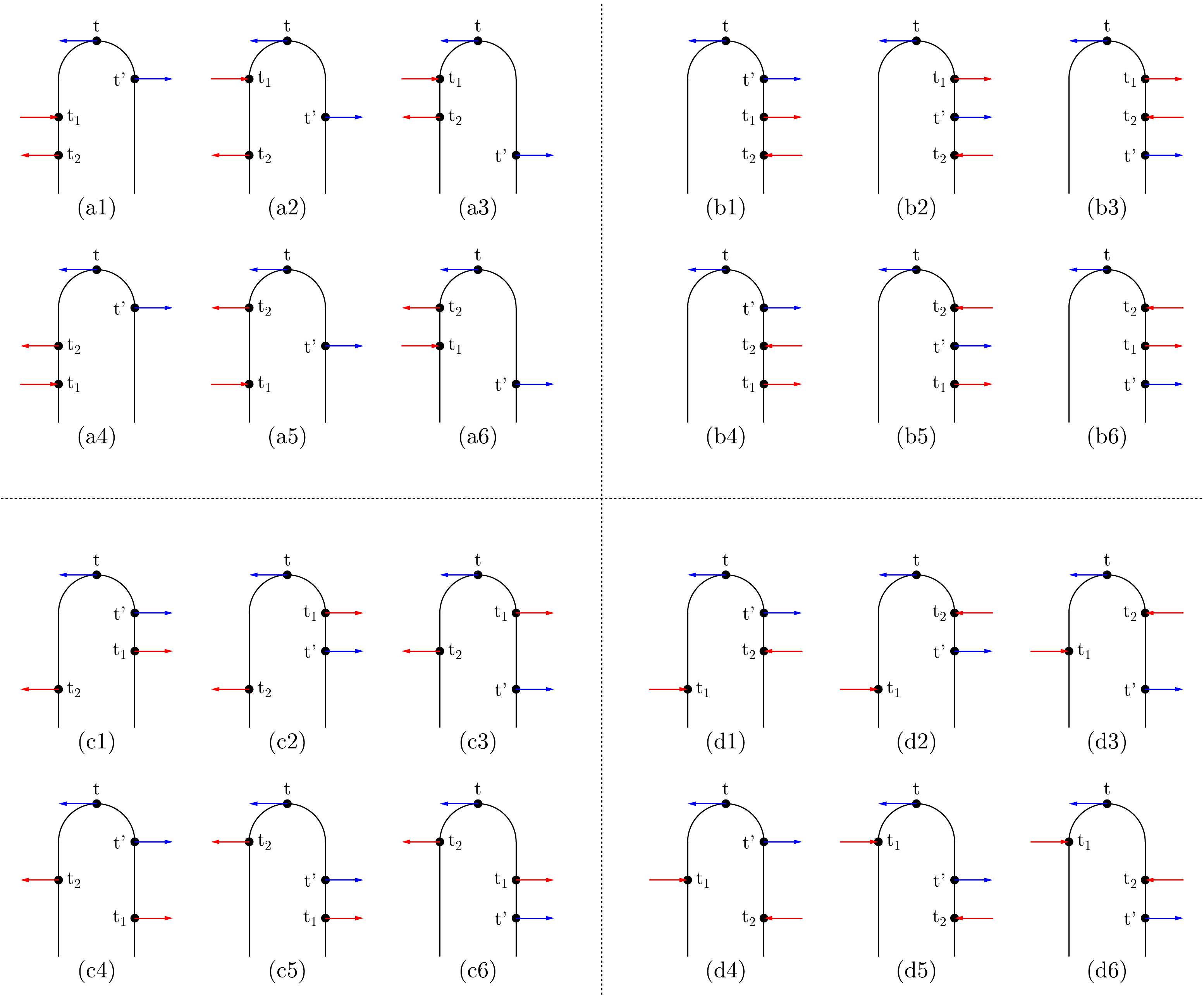}
\caption{\label{fig3}
Contour projections (double-sided Feynman diagrams) for the fourth order interaction 
with the modes $i$ (red, times $t_1$ and $t_2$) and $f$ (blue, times $t$ and $t'$) 
of the optical field. Time increases from bottom to top.
}
\end{figure}

The Liouville space procedure applied to Eq.~(\ref{Jif})
deals with the same problem of ordering variables $\tau_1$ and $\tau_2$
between the two times $t$ and $t'$. However this time the ordering
is performed along the real time axis (i.e. not only relative position 
of times on the contour but also relative position on the real time axis 
is tracked), thus number of diagrams (different orderings) is bigger here.
Information on the branch of the contour is provided by 
the two-side Feynman diagrams (the Keldysh contour diagrams
with additional restriction on the relative positions of times 
with respect to the real time axis). It is customary to indicate
each photon process by separate arrow in these diagrams, rather 
than consider contractions representing free photon propagation.
The agreement is that arrow pointing to the left corresponds to
creation operator of the photon in quantum mechanical description of the field
(or factor $e^{i\nu t}$ for classical treatment of the field), 
while arrow pointing to the right represents operator of annihilation
of the photon (or factor $e^{-i\nu t}$)~\cite{Mukamel_1995}.  
The procedure being applied to expression (\ref{Jif}) leads to $24$
double-sided Feynman diagrams presented in Fig.~\ref{fig3}.
These are the diagrams presented in Figs.~4 and 5 of Ref.~\onlinecite{HarbolaMukamelJCP14}.
Similar to the Hilbert space formulation $24$ more diagrams are complex
conjugate variants.  

\begin{figure}[t]
\centering\includegraphics[width=\linewidth]{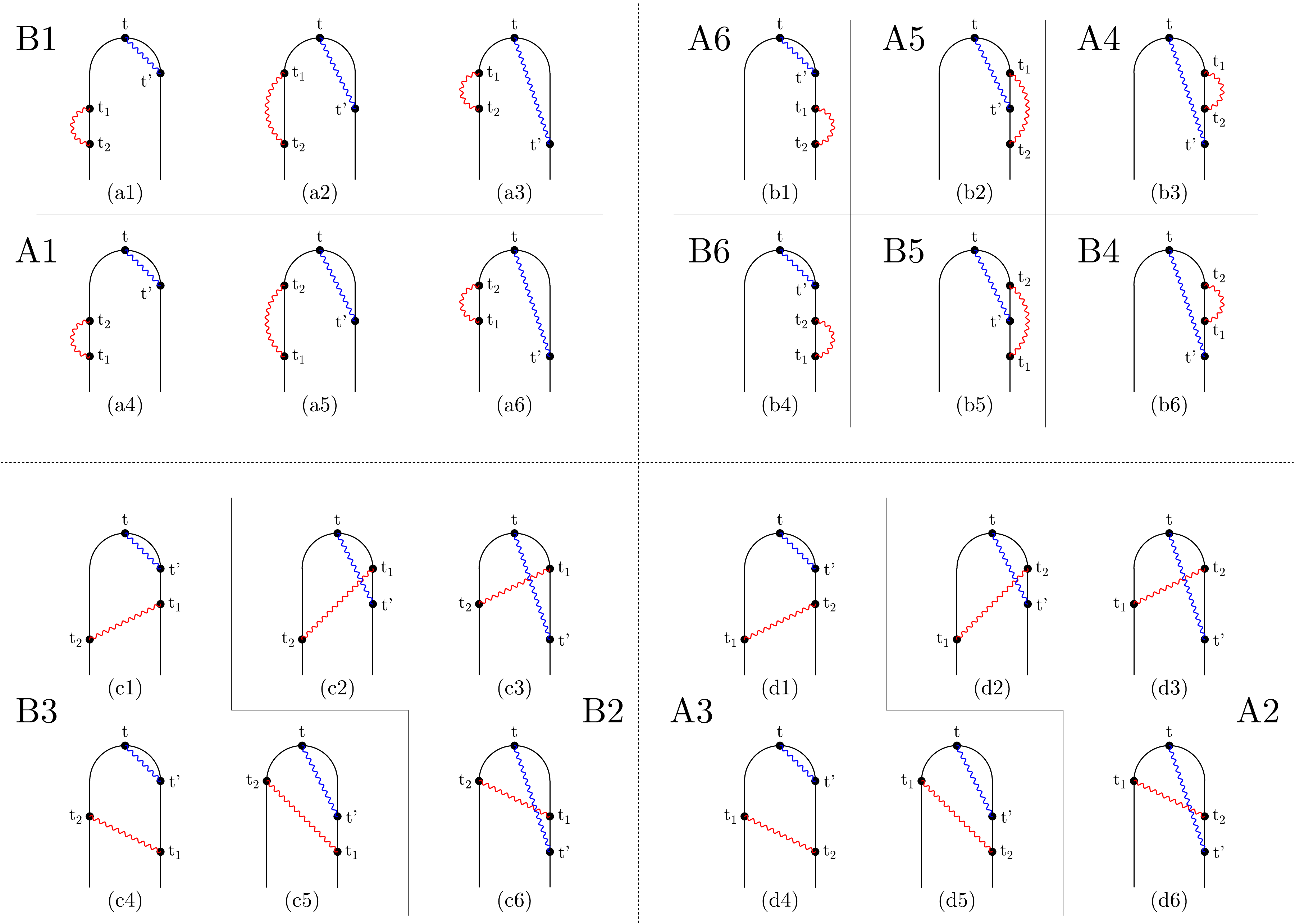}
\caption{\label{fig4}
Comparison between contour projections presented within the Hilbert (see Fig.~\ref{fig2})
and Liouville (see Fig.~\ref{fig3}) space approaches. Time increases from bottom to top.
}
\end{figure}

Comparison of the two sets of diagrams is straightforward, when one
keeps in mind an additional restriction on time ordering within
the Liouville space formulation (ordering not only on the contour but also
with respect to real time axis). So, one Hilbert space type diagram
(see Fig.~\ref{fig2}) may encompass several diagrams of the Liouville type
(see Fig.~\ref{fig3}). Diagram-to-diagram correspondence is shown
in Fig.~\ref{fig4}.


\section{Quasiparticles vs. many-body states approach}\label{qp_mb}
In realistic simulations in junctions in addition to molecular
interaction with radiation field other intra-molecular interactions 
(e.g. electron-electron or electron-phonon) and couplings to 
environment (e.g. coupling to contacts) should be accounted for.

Quasiparticles approaches (in either Hilbert or Liouville space) 
are usually efficient in treating molecule-contacts couplings.
Indeed, since the latter are represented as quadratic in terms of 
elementary excitations, Eq.~(\ref{VSB}), they do not alter non-interacting character 
of the Hamiltonian, and thus Wick's theorem is available to
exactly evaluate multi-time correlation functions in presence
of the couplings~\cite{Danielewicz1984}.  
However, quasiparticles are less convenient when dealing with
intra-molecular interactions. That is, if intra-molecular interactions 
are weak compared to either molecule-contacts couplings or separation 
between electron energy from  molecular resonances, 
standard diagrammatic technique (on the contour perturbation theory) 
can be invoked to account approximately for the interactions.

For strong interactions (e.g. Coulomb blockade or polaron formation) 
usually encountered in the resonant tunneling regime, a nonequilibirum
atomic limit (i.e. consideration utilizing many-body states
of the isolated molecule as a basis) is preferable~\cite{WhiteOchoaGalperinJPCC14}.
Note that historically spectroscopy of isolated molecules 
also was formulated mostly in the language of many-body states
(molecular or dressed states of molecule and the field)~\cite{Mukamel_1995}.
Many-body states based formulations, 
while accounting exactly for intra-molecular interactions, 
are capable of treating molecule-contacts couplings only approximately.
We note that such approximate schemes exist for both 
Hilbert~\cite{SandalovIJQC03,FranssonPRB05,GalperinNitzanRatnerPRB08}
and Liouville~\cite{LeijnseWegewijsPRB08,EspGalpPRB09,WegewijsPRB12}
space formulations, however these approximations are not well controlled.

A promising tool is the Hilbert space type pseudoparticle NEGF (PP-NEGF)
technique~\cite{WernerPRB13,OhAhnBubanjaPRB11,WhiteGalperinPCCP12,WernerRMP14},
which while being a many-body states formulation allows to
take into account coupling to the contacts within well controllable
diagrammatic technique, which is based on extension of the standard methods
of quantum field theory into extended version of the Hilbert space.
The PP-NEGF has several important advantages: 
1. The method is conceptually simple; 
2. Its practical implementations rely on a set of controlled approximations 
(standard diagrammatic perturbation theory techniques can be applied); 
3. Already in its simplest implementation, the non-crossing approximation
(NCA), the pseudoparticle NEGF goes beyond standard
QME approaches by accounting for both non-Markovian
effects and hybridization of molecular states; 
4. The method is capable of treating junction problems in the language of many-body
states of the isolated molecule, exactly accounting for all the
on-the-molecule interactions. 
Recently a first application of the methodology to problems
of optical spectroscopy of junctions was proposed in Ref.~\onlinecite{WhiteTretiakNL14}.


\section{Raman spectroscopy}\label{raman}
Eq.~(\ref{Jif}) represents all contributions to the fourth order optical process.
Note that interaction with mode $f$ only includes outgoing photon, while
interactions with mode $i$ have both incoming and outgoing photon contributions.
Correspondingly, the projections can be separated into those containing
two outgoing photons (diagrams B in Fig.~\ref{fig2}) and those which contain
incoming photon of mode $i$ and outgoing photon of mode $f$ 
(diagrams A in Fig.~\ref{fig2}). The former contribute to fluorescence 
(both coherent fourth order and sequential second order processes),
while the latter yield Raman (see below) and absorption with emission
(sequential second order) processes. 

Both diagrams A and B of Fig.~\ref{fig2} contain virtual photon
processes. For example, interaction with mode $i$ in projections (a4) (A type diagram) 
and (a1) (B type diagram) of Fig.~\ref{fig3} are virtual photon processes. 
These processes yield renormalization of molecular correlation functions 
due to presence of the radiation field.
Accounting for such processes within the procedure described above should be done 
with care. Indeed, reasonable estimate of the interaction with the field is 
$U\sim 10^{-3}-10^{-2}$~eV~\cite{SukharevPRB10}.
For a molecule chemisorbed on metallic surface electronic escape rate
(characteristic strength of the molecule-contact 
interaction) is  $\Gamma\sim 0.01-0.1$~eV~\cite{KinoshitaJCP95}, and
in this case renormalizations due to coupling to contacts are much more pronounced.
For a molecule detached from contacts renormalizations due to the radiation field
become important, however proper way to account for those is by solving 
Eq.~(\ref{Dyson}). Accouting for the renormalization by keeping only second term
in the right of Eq.~(\ref{PT}) (besides being a non-conserving approximation)
results in a Green function divergent at molecular resonances.
Resummation of the perturbation series and introduction of the retarded projection of 
the self-energy is the proper way to account for the renormalization, when
coupling to the radiation field is weak relative to intra-molecular interactions.
If coupling to the radiation field is the dominating interaction, perturbation series
treatment becomes invalid.
We stress that in experiment one deals with the total optical signal, and a
proper theoretical treatment should utilize Eq.~(\ref{flux}) with conserving 
approximations employed to evaluate molecular Green functions. 

We now focus on theoretical description of a particular optical process 
- the Raman scattering. 
Raman scattering from mode $i$ to mode $f$ is a coherent process of (at least) 
fourth order in coupling to radiation field with two orders in coupling to incoming photon 
of mode $i$ and two orders in coupling to outgoing photon of mode $f$. 
The process should also satisfy a proper energy conservation:
at steady-state difference between incoming and outgoing photon frequencies should 
be equal to multiples of the frequency of molecular vibration (may be plus electronic energy
differences within the same electronic level or many-body state broadened by 
coupling to environment)
\begin{equation}
\label{energy}
J_{i\to f}^{Raman}\sim \delta(\nu_i-\nu_f+n\omega_v+\Delta E)
\end{equation}
Here $n$ is integer number ($n=0$ yields Rayleigh scattering) 
and $\Delta E$ is electronic energy change
within the same broadened level or many-body state.

Identification of diagramms related to Raman scattering requires some care. 
An easy way to identify relevant contributions relies on the Hilbert space 
formulation of section~\ref{hilbert_liouv} with utilization of the Langreth projection
rules~\cite{HaugJauho_2008}.
First, only projections A of Fig.~\ref{fig2} are those corresponding to a process with one 
incoming and one outgoing photon. Second, among those only parts of projections 
A2, A3, A4, and A5 yield proper energy conservation, Eq.~(\ref{energy}).
For example, let employ the Langreth rules to projection A4. This yields 
4 terms in expression (\ref{Jif})
\begin{subequations}
\label{integral}
\begin{align}
& \int_{-\infty}^{+\infty} d(t'-t)\int_{-\infty}^0 d(t_1-t)\int_{-\infty}^{0} d(t_2-t_1)\ldots 
\\ 
\label{eqRaman}
-&
 \int_{-\infty}^{+\infty} d(t'-t)\int_{-\infty}^0 d(t_1-t)\int_{-\infty}^{0} d(t_2-t')\ldots  \\
 -&
\int_{-\infty}^{+\infty} d(t'-t)\int_{-\infty}^0 d(t_1-t')\int_{-\infty}^{0} d(t_2-t_1)\ldots
\\ 
+&
 \int_{-\infty}^{+\infty} d(t'-t)\int_{-\infty}^0 d(t_1-t')\int_{-\infty}^{0} d(t_2-t')\ldots 
\end{align}
\end{subequations}
Only (\ref{eqRaman}) yields proper energy conservation.
Similarly, application of the Langreth rules to projection A3 yields one, 
and to projections A2 and A5 yields two integrals for each with only one term in 
each case satisfying restrictions of the Raman process. 

In Refs.~\onlinecite{GalperinRatnerNitzanNL09,GalperinRatnerNitzanJCP09}
we presented a theory of Raman scattering in molecular junctions
formulated within the NEGF (the Hilbert space quasiparticle formulation).
Projections presented there were different from those given in Fig.~\ref{fig2}
in two aspects: a.~the projections there are given after the Langreth rules
have been applied and b.~some of the projections (see e.g. Fig.~8b of Ref.~\onlinecite{GalperinRatnerNitzanJCP09})
represent `the hole view' of scattering process. 
We note that in general (when explicit time-dependent processes or a magnetic field 
are present) the latter flexibility does not exist. 
However for steady-state situation the two pictures can be shown to be equivalent.
For example, the contribution (\ref{eqRaman}) to the Raman signal is
(note, it is easy to see that the contribution is real following the derivation leading form Eq.(28) to Eq.(56) in Ref.~\onlinecite{GalperinRatnerNitzanJCP09})
\begin{align*}
\label{JifA4}
& \vert U_i\, U_f\rvert^2 N_i
 \int_{-\infty}^{+\infty} d(t'-t) \int_{-\infty}^0 d(t_1-t)\int_{-\infty}^0 d(t_2-t')\,
 e^{i\nu_f(t-t')-i\nu_i(t_1-t_2)}
 \\ &\qquad\qquad\qquad \times
\langle \hat O^\dagger(t')\,\hat O(t_2)\,\hat O^\dagger(t_1)\,\hat O(t)\rangle 
\\ &
\equiv\lvert U_i\, U_f\rvert^2 N_i
 \int_{-\infty}^{+\infty} d(t'-t) \int_{-\infty}^0 d(t_1-t)\int_{-\infty}^0 d(t_2-t')\,
 e^{-i\nu_f(t-t')}\,e^{i\nu_i(t_1-t_2)}
 \\ &\qquad\qquad\qquad \times
\langle \hat O^\dagger(t)\,\hat O(t_1)\,\hat O^\dagger(t_2)\,\hat O(t')\rangle
\end{align*}
where $N_i$ is the population of the mode $i$.
Then utilizing the time-reversal symmetry of the molecular correlation function,\cite{Banyai_2006}
\begin{equation}
 \langle \hat O^\dagger(t)\,\hat O(t_1)\,\hat O^\dagger(t_2)\,\hat O(t')\rangle
= 
\langle \hat O^\dagger(-t')\,\hat O(-t_2)\,\hat O^\dagger(-t_1)\,\hat O(-t)\rangle
\end{equation}
and inverting signs of the time variables
one gets Eq.~(29) of Ref.~\onlinecite{GalperinRatnerNitzanJCP09},
which was derived from `the hole type' projection shown in Fig.~8b there.
We note that while the results are equivalent, graphical forms of the projections differ: 
Fig.~8b of Ref.~\onlinecite{GalperinRatnerNitzanJCP09}
has the two incoming field modes on different branches,  
while the diagram A4 - on the same branch.


\section{Conclusion}\label{conclude}
We discussed theoretical formulations utilized in studies of 
optical processes in current carrying  molecular junctions. 
We start by identifying total optical signal as photon flux from the system to registering device,
for which exact formulation in terms of molecular correlation (Green) functions is available.  
Standard quantum field theory methods allow to formulate well controlled conserving
approximation to evaluate the Green functions of the molecule. 
At the same time tradition coming from the theory of optical spectroscopy
for isolated molecules utilizes perturbation theory.
The latter, while not being a conserving approximation,  
allows to separate different photon scattering events in the total
optical signal. 
We discussed the Hilbert and Liouville space formulations,
indicating similarities and differences in graphical representations
of photon scattering process in the two formulations. 
We also pointed out strong and weak sides of quasiparticles (second quantization) 
vs. many-body states based  (nonequilibrium atomic limit) approaches in both formulations, 
and outlined the PP-NEGF as a promising theoretical approach to study optical 
spectroscopy in junctions. 
Finally, we focused on theoretical formulation for Raman scattering
in molecular junctions, and clarified questions raised in 
the literature~\cite{HarbolaMukamelJCP14} with respect
to identification of relevant projections and obtaining corresponding contributions
to the total optical signal. Thus, the paper bridges
different formulations of optical spectroscopy in open nonequilibrium systems.


%
\end{document}